\documentclass[pra,aps,twocolumn,showkeys,showpacs,superscriptaddress]{revtex4-1}

\usepackage{graphicx}
\usepackage{amsmath}
\usepackage{amssymb}
\usepackage{bm}
\usepackage{dsfont}
\usepackage{graphicx}
\usepackage{subfigure}
\usepackage{times}
\usepackage{url}
\usepackage[colorlinks=true,citecolor=blue,urlcolor=blue]{hyperref}
\usepackage[mathscr]{euscript}

\usepackage[x11names,dvipsnames]{xcolor}
\definecolor{vured}{RGB}{137,28,46}
\definecolor{vugrey}{RGB}{197,197,197}
\definecolor{vugreen}{RGB}{34,139,34}
\definecolor{midnightblue}{RGB}{25,25,112}

\newcommand{\myblue}{\color{midnightblue}}

\newcommand{\p}{\partial}
\newcommand{\rA}{\mathcal{A}}
\newcommand{\rB}{\mathcal{B}}
\newcommand{\rD}{\mathrm{D}}
\renewcommand{\ao}{\hat{a}}
\renewcommand{\aa}{\hat{a}^{\dag}}
\newcommand{\bo}{\hat{b}}
\newcommand{\ba}{\hat{b}^{\dag}}
\newcommand{\Ho}{\hat{H}}
\newcommand{\la}{\langle}
\newcommand{\ra}{\rangle}
\newcommand{\re}{\mathrm{e}}
\newcommand{\ri}{\mathrm{i}}
\newcommand{\ph}{\varphi}

\newcommand{\rJ}{\mathcal{J}}
\newcommand{\bk}{\bm{k}}
\newcommand{\br}{\bm{r}}
\newcommand{\bF}{\bm{F}}
\newcommand{\bma}{\bm{a}}
\newcommand{\bmb}{\bm{b}}
\newcommand{\Uo}{\hat{U}}
\newcommand{\Ua}{\hat{U}^{\dag}}
\newcommand{\Vo}{\hat{V}}
\newcommand{\inter}{\text{int}}
\newcommand{\no}{\hat{n}}

\begin{document}

\title{\myblue Modified interactions in a Floquet topological system on a square 
lattice \\ and their impact on a bosonic fractional Chern insulator state}
\date{\today}

\author{Mantas Ra\v{c}i\={u}nas}
\affiliation{Institute of Theoretical Physics and Astronomy, Vilnius University,
A. Go\v{s}tauto 12, LT-01108 Vilnius, Lithuania}
\author{Giedrius \v{Z}labys}
\affiliation{Institute of Theoretical Physics and Astronomy, Vilnius University,
A. Go\v{s}tauto 12, LT-01108 Vilnius, Lithuania}
\author{Andr\'e Eckardt}
\email{eckardt@pks.mpg.de}
\affiliation{Max-Planck-Institut f\"ur Physik komplexer Systeme, 
\mbox{N\"othnitzer Stra\ss e 38, 01187 Dresden, Germany}}
\author{Egidijus Anisimovas}
\email{egidijus.anisimovas@ff.vu.lt}
\affiliation{Institute of Theoretical Physics and Astronomy, Vilnius University,
A. Go\v{s}tauto 12, LT-01108 Vilnius, Lithuania}

\begin{abstract}
We propose a simple scheme for the realization of a topological quasienergy band structure 
with ultracold atoms in a periodically driven optical square lattice. It is based on a 
circular lattice shaking in the presence of a superlattice that lowers the energy on every 
other site. The topological band gap, which separates the two bands with Chern numbers 
$\pm 1$, is opened in a way characteristic to Floquet topological insulators, namely, by 
terms of the effective Hamiltonian that appear in subleading order of a high-frequency 
expansion. These terms correspond to processes where a particle tunnels several times 
during one driving period. The interplay of such processes with particle interactions also 
gives rise to new interaction terms of several distinct types. For bosonic atoms with 
on-site interactions, they include nearest neighbor density-density interactions introduced 
at the cost of weakened on-site repulsion as well as density-assisted tunneling. Using 
exact diagonalization, we investigate the impact of the individual induced interaction terms 
on the stability of a bosonic fractional Chern insulator state at half filling of the lowest 
band.
\end{abstract}


\maketitle

\section{Introduction}

The powerful concept of Floquet engineering
\cite{goldman14,eckardt15,goldman15resonant,bukov15,holthaus16} 
is based on the possibility to emulate effective time-independent Hamiltonians with desired 
properties by subjecting a suitably chosen and well controllable physical system to a 
time-periodic external field. This procedure relies on the observation that the evolution 
of a quantum system described by a time-periodic Hamiltonian is governed by a 
time-independent effective Hamiltonian 
\cite{rahav03,goldman14,eckardt15,goldman15resonant,bukov15}. 
It is particularly relevant for modern experiments on cold-atom systems in optical lattices
\cite{bloch08,dalibard11,lewenstein12,windpassinger13,goldman14rpp},
which are prominent due to their unsurpassed level of control, nearly perfect structural 
purity as well as the high degree of isolation and consequent minimal dissipation. 
Complementing these advantages with a specific driving protocol that allows for a clear 
physical interpretation of the resulting effective Hamiltonian (computed within a 
suitable approximation), one is able to carry out quantum simulation of paradigmatic 
physical models and realize novel or elusive phases of matter. 
Successful experiments include the demonstration of the basic quantum phase transition 
between superfluid and Mott-insulating phases \cite{eckardt05,zenesini09}, emulation of 
spin models \cite{eckardt10,struck11,struck13},
realization of intense artificial magnetic fields
\cite{aidelsburger11,bermudez11,kolovsky11,aidelsburger13,aidelsburger15,%
struck12,struck13,miyake13,atala14,kennedy15}, 
and topological band structures \cite{jotzu14,aidelsburger15}.
In addition to the detection of integrated topological characteristics (Chern numbers),
full tomography of the Berry curvature \cite{hauke13} has recently been 
achieved \cite{flaeschner15}.

Concerning topological band structure engineering, it is important to make a clear 
distinction between ``fast'' and ``intermediate'' driving schemes. Here the relevant 
parameters are the energy scale $\hbar\omega$ defined by the driving frequency $\omega$ 
and the characteristic energy of the system's internal degrees of freedom; for lattice 
systems, typically the hopping parameter $J$. The limit of rapid forcing is defined by 
the condition $\hbar\omega \gg J$, and the effective Hamiltonian is adequately approximated
by the straightforward time average of the driven Hamiltonian. A number of schemes 
of this type were proposed and realized
\cite{eckardt05,lignier07,sias08,zenesini09,alberti09,eckardt10,haller10,struck11,%
kolovsky11,aidelsburger11,hauke12,struck12,struck13,aidelsburger13,miyake13,jotzu14,%
atala14,creffield14,aidelsburger15,kennedy15,flaeschner15}. 
Going beyond the time-averaging of the driven Hamiltonian one relies on an expansion 
in powers of the inverse frequency (or, equivalently, the period)
\cite{goldman14,eckardt15,goldman15resonant,bukov15,mikami15}
and includes successive terms proportional to, e.~g., $J/\hbar\omega$ and $(J/\hbar\omega)^2$. 
Schemes relying on these contributions have been termed Floquet topological insulators
\cite{oka09,kitagawa10,lindner11,kitagawa11,cayssol13,grushin14}
and have been demonstrated not only in an optical lattice \cite{jotzu14}, but also in 
photonic wave guides \cite{rechtsman13}. These subleading terms have a clear physical 
interpretation \cite{eckardt15,anisimovas15}
which is an essential ingredient that makes the term-by-term construction of the effective 
Hamiltonians meaningful. The relevant corrections are: (a) processes where a particle tunnels 
twice during a driving period thus generating effective matrix elements for tunneling beyond 
nearest neighbors, and (b) combined events involving an interplay between kinetics and 
interactions, which are the main focus of the present manuscript. Both types of processes are 
intimately related to the presence of a significant micromotion corresponding to the periodic 
motion of particles in real space at the driving frequency. In a recent study 
\cite{anisimovas15}, the coupling of micromotion and interactions was shown to be largely 
detrimental to the stabilization of the fractional Chern insulator phases
\cite{sheng11,neupert11,sun11,regnault11,wang11,wu12,grushin14,grushin15stability,%
bergholtz13,parameswaran13}. However, a detailed analysis and an insight into the 
underlying mechanism was not given. 

The aim of the present paper is twofold: On the one hand we propose a scheme for the 
realization of a Floquet topological band structure with two Chern bands 
\footnote{In our work we exclusively focus on Chern insulators \cite{haldane88} belonging 
to the basic class A of more general topological insulators \cite{hasan10,qi11}.} 
in a circularly driven square lattice. The scheme  reproduces the physics of the chiral 
$\pi$-flux model \cite{neupert11,sun11,regnault11} and relies on engineering the 
necessary flux configuration by ``photon''-assisted hoppings in the presence of 
sublattice modulation, while the topological band gaps are opened due to induced 
next-neareast neighbor transitions. On the other hand, we investigate the stability 
of the fractional Chern insulator phase \cite{bergholtz13,parameswaran13} of bosonic 
particles in the half-filled lowest energy band. In particular, we focus on the impact 
of different micromotion-induced interaction terms \cite{eckardt15,anisimovas15} in the 
effective Hamiltonian, investigating in detail which of these terms are beneficial and 
which detrimental for the preparation of the fractional Chern insulator state.

We find that micromotion-induced corrections to particle interactions can be separated 
into three constituent components: (i) weakening of the on-site interaction strength, 
(ii) the appearance of induced interactions between neighboring sites even if they were 
absent in the original model, and (iii) the remaining density-assisted tunneling events.
Interestingly, contributions of density-density interaction type, (i) and (ii), satisfy 
a constraint in the form of a sum rule indicating that the diminished on-site interaction 
energy is precisely compensated by the corresponding increase of nearest-neighbor 
interaction energies; in other words, the interactions are ``smeared out'' by micromotion. 
We study the impact of the three effects on the stability of the fractional Chern insulator 
phase and demonstrate that, as the driving frequency becomes lower, this phase is primarily
destabilized by the destructive role of the density-assisted hopping terms.

Our paper has the following structure. In Sec.~\ref{sec:model}, we present a feasible 
scheme of practical realization of the chiral $\pi$-flux model supporting robust 
topological single-particle energy bands. The inclusion of inter-particle interactions 
and their coupling to the real-space micromotion are discussed in Sec.~\ref{sec:inter} 
and supported with numerical results in Sec.~\ref{sec:numpy}. We summarize our findings 
in the concluding Sec.~\ref{sec:sum}, while a number of issues of technical nature are 
delegated to Appendices. The topics include the creation of artificial gauge structures, 
high-frequency expansion of the effective Hamiltonian, and supplemental analysis of 
single-particle band structures.

\section{Modulated square lattice}
\label{sec:model}

In this section, we describe a specific driving scheme that allows to realize the chiral 
$\pi$-flux model \cite{neupert11,sun11,regnault11} on a modulated square lattice. As the 
name implies, this model features elementary plaquettes pierced by (artificial) magnetic 
fluxes equal to one half of the dimensionless flux quantum $2\pi$. Alongside with the 
Haldane model \cite{haldane88,alba11,goldman13} based on a hexagonal 
lattice, the chiral $\pi$-flux model presents an unpretentious two-band configuration that 
serves as a basis for robust topological band structures. In the presence of only 
nearest-neighbor hopping both models support band structures featuring two Dirac points 
where the upper and the lower bands touch at a singular point in a cone-like fashion. 
Inclusion of next-nearest neighbor transitions leads to the opening of topological band 
gaps whereby the two energy bands separate and may acquire the Chern indices of $\pm 1$. 
The presence of additional tunable parameters (essentially, the ratio of nearest and 
next-nearest neighbor transition strengths) allows one to tune into the regime where 
topological bands are relatively flat in comparison to band gaps and interaction energies. 
These circumstances pave the wave towards the realization of so far elusive fractional 
Chern insulators.

\subsection{\myblue Driven Hamiltonian and flux configuration}

\begin{figure}
\includegraphics[width=84mm]{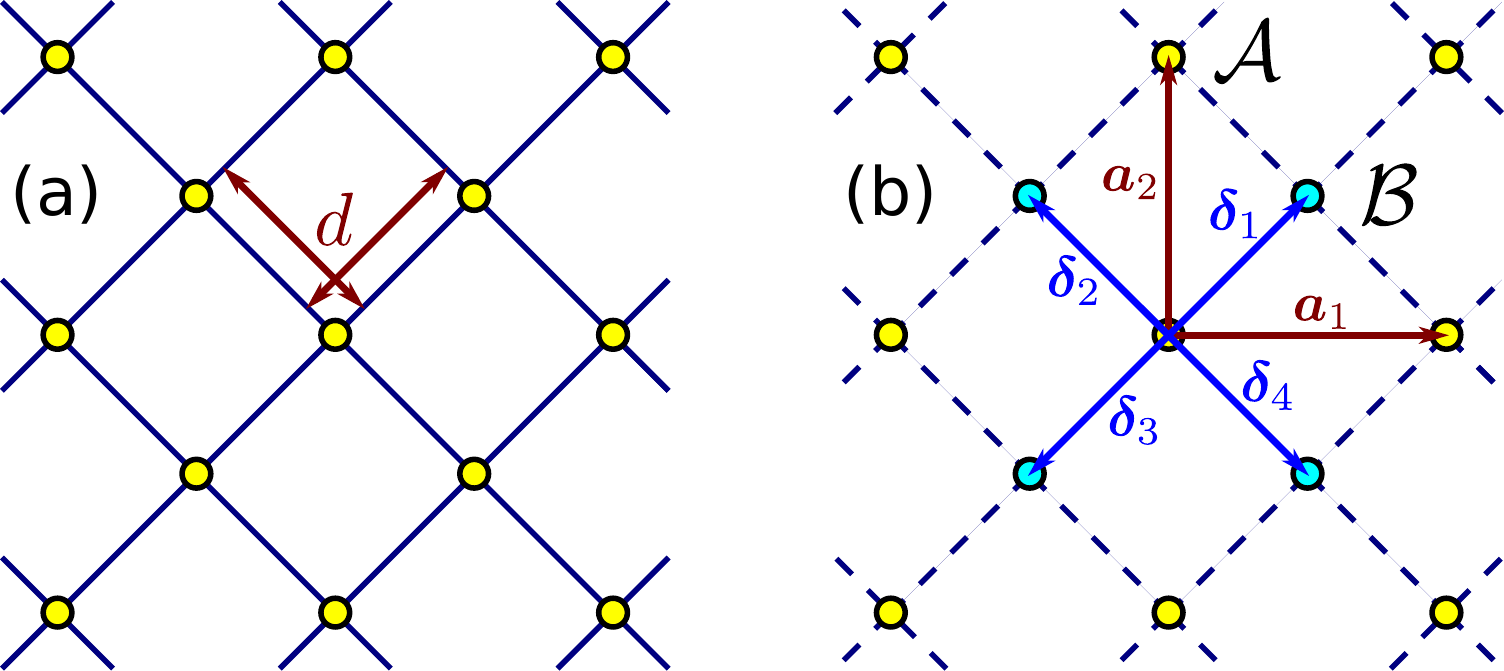}
\caption{\label{fig:latt} Realizing the chiral-$\pi$ model. The original square lattice
in panel (a) is separated into two energy-mismatched sublattices (yellow $\rA$ and cyan 
$\rB$) of a larger lattice constant in panel (b). The vectors $\bm{\delta}_{\mu}$ with 
$\mu = \{1, 2, 3, 4 \}$ connect the nearest-neighbor sites belonging to distinct 
sublattices. The vectors $\bm{a}_{1|2}$ are the elementary translation vectors. Full 
(dashed) blue lines depict natural (driving-assisted) transitions.}
\end{figure}

In order to produce the required flux configuration, we start from a static square 
lattice of spacing $d$ shown in Fig.~\ref{fig:latt}~(a). Here, the full blue lines denote 
nearest-neighbor links characterized by a spatially uniform bare tunneling parameter $J$. 
Next, we bipartition the original lattice into two square sublattices $\rA$ and $\rB$ with 
lattice constants $\sqrt{2}d$, intertwined in a checkerboard fashion. By means of lowering 
the on-site energies on sublattice $\rB$ by the quantity $\hbar\omega$, the natural 
hopping transitions between nearest-neighbor sites are inhibited and must be assisted by 
an external driving at frequency $\omega$. The resulting lattice configuration is shown 
in Fig.~\ref{fig:latt}~(b) with dashed blue lines indicating such ``photon''-assisted 
transitions.

The effects of a time-periodic lattice forcing are captured by the driven single-particle
Hamiltonian 
\begin{equation}
\label{eq:one}
  \Ho (t) = - J \sum_{\la ij \ra} \aa_i \ao_j 
  + \sum_{i} v_i (t) \aa_i \ao_i,
\end{equation}
with the lattice degrees of freedom encoded by annihilation (creation) operators 
$\ao_i^{(\dag)}$. The Hamiltonian (\ref{eq:one}) is composed of the static part representing 
transitions on all directed lattice links $\la ij \ra$ plus the driving term featuring time- 
and coordinate-dependent on-site potentials $v_i(t)$. In the case of circular driving by 
a rotating force of magnitude $F$ and on-site energy offsets $(-\hbar\omega)$ affecting only 
sublattice $\rB$, a further gauge transformation (see Appendix~\ref{app:gauge} for details) 
leads to the purely kinetic driven Hamiltonian in the form 
\begin{equation}
\label{eq:hot1}
  \Ho (t) = - \sum_{i \in \rA} \sum_{\mu = 1}^{4} J_{\mu} (t) \, 
  \ba_{i+\mu} \ao_i + \mathrm{h.c.},
\end{equation}
with $\ao_i^{(\dag)}$ and $\bo_i^{(\dag)}$ denoting the annihilation (creation)
operators defined on the respective sublattices. In order to count all nearest-neighbor 
connections only once, we sum over all sites belonging to the sublattice $\rA$ and over 
the four distinct directions $\bm{\delta}_{\mu}$ (see  Fig.~\ref{fig:latt} for 
definitions) linking to the nearest neighbors belonging to the sublattice $\rB$. 
The time-dependent hopping parameters are given by [cf.~(\ref{eq:aten})]
\begin{equation}
\label{eq:hopa}
  J_{\mu} (t) = J \, \re^{\ri\alpha\sin(\omega t - \ph_{\mu})} \re^{-\ri\omega t},
\end{equation}
with the scaled shaking strength $\alpha = Fd / \hbar\omega$ and the lagging phases
\begin{equation}
\label{eq:lag}
  \ph_{\mu} = \frac{\pi (\mu-1)}{2}
  = \Big\{ 0, \frac{\pi}{2}, \pi, \frac{3\pi}{2} \Big\}.
\end{equation}
The presence of the exponential factor $\re^{-\ri\omega t}$ in the hopping parameter 
(\ref{eq:hopa}) results from the sublattice energy mismatch. We see that the time 
dependence enters the driven Hamiltonian (\ref{eq:hot1}) only through the modulation 
of the hopping parameters whose Fourier series read
\begin{equation}
\label{eq:jaymut}
  J_{\mu} (t) = \sum_{s=-\infty}^{\infty} 
  J \rJ_{1+s} (\alpha) \, \re^{-\ri(1+s)\ph_{\mu}}
  \, \re^{\ri s \omega t},
\end{equation}
with $\rJ_m (x)$ denoting the Bessel function of the first kind and order $m$. 
In the high-frequency limit, when time averaging of the driven Hamiltonian is an 
adequate approximation, one has
\begin{equation}
  \langle J_{\mu} (t) \rangle_T = J \rJ_1 (\alpha) \, \re^{-\ri\ph_{\mu}},
\end{equation}
so that the Peierls phases associated with transitions from sublattice $\rA$ to sublattice 
$\rB$ are given by the lagging phases (\ref{eq:lag}). Keeping in mind that transitions
in the opposite direction are associated with complex conjugate matrix elements and
inverted phases, it is easy to check that the total phase accumulated while travelling 
around each square plaquette equals 
\[
  - \frac{\pi}{4} + \frac{3\pi}{4} - \frac{5\pi}{4} + \frac{7\pi}{4} = \pi.
\]

\subsection{\myblue Description in quasimomentum space}

To proceed with further analysis, it is convenient to switch into the reciprocal space, 
which is accomplished through the introduction of the quasimomentum-dependent operators
\begin{subequations}
\begin{align}
  \aa_{\bk} 
    &= \frac{1}{\sqrt{N_s}} \sum_{i \in \rA} \aa_i \, 
    \re^{\ri \bk\cdot\br_i}, \\
  \ba_{\bk} 
    &= \frac{1}{\sqrt{N_s}} \sum_{j \in \rB} \ba_j \, 
    \re^{\ri \bk\cdot (\br_j - \bm{\delta}_1)}.
\end{align}
\end{subequations}
Here $N_s$ is the number of sites in a given sublattice.
On the second line, the sum runs over all sites $j$ belonging to the sublattice $\rB$, 
and the vector $\br_j$ points to the position of a particular site. The additional shift 
of its effective position by $\bm{\delta}_1$ is included to obtain a $\bk$-periodic
Hamiltonian \cite{regnault11}. Since the driven Hamiltonian (\ref{eq:hot1}) only couples 
sites belonging to different sublattices, we find
\begin{equation}
\label{eq:driven1}
  \Ho (t) = - J \sum_{\bk} \ba_{\bk} \ao_{\bk} \, g (t,\bk)
    - J \sum_{\bk} \aa_{\bk} \bo_{\bk} \, g^*(t,\bk),
\end{equation}
with the matrix element $g (t, \bk)$ encompassing the summation over the four 
nearest-neighbor links
\begin{equation}
\begin{split}
\label{eq:driven2}
  g (t,\bk) &= \sum_{\mu=1}^4 \left[J_{\mu} (t)/J \right] 
   \, \re^{-\ri\bk\cdot(\bm{\delta}_{\mu} - \bm{\delta}_1)} \\
  &= \sum_{s=-\infty}^{\infty} \rJ_{1+s} (\alpha) G_{1+s} (\bk) \, \re^{\ri s\omega t}.
\end{split}
\end{equation}
Here, we found it convenient to introduce a family of auxiliary functions
\begin{equation}
\begin{split}
\label{eq:driven3}
  G_s (\bk) &= 1 
    + \re^{-\ri s \pi/2} \re^{\ri\bk\cdot\bma_1} \\
  &+ \re^{\ri s \pi} \re^{\ri\bk\cdot(\bma_1+\bma_2)} 
    + \re^{\ri s \pi/2} \re^{\ri\bk\cdot\bma_2},
\end{split}
\end{equation}
whose properties are analyzed in Appendix~\ref{app:fungus}. Finally, the 
operator-valued Fourier coefficients of the driven kinetic Hamiltonian are given by
\begin{equation}
\begin{split}
\label{eq:try}
  \Ho_s = &-J \rJ_{1+s} (\alpha) \sum_{\bk} \ba_{\bk} \ao_{\bk} G_{1+s} (\bk) \\
  &-J \rJ_{1-s} (\alpha) \sum_{\bk} \aa_{\bk} \bo_{\bk} G_{1-s}^* (\bk).
\end{split}
\end{equation}

\subsection{\myblue Effective Hamiltonian}

For a periodically driven system, the quantum-mechanical time-evolution operator between
arbitrary times $t_1$ and $t_2$ can be factorized like (see, e.~g., 
\cite{goldman14,goldman15resonant,eckardt15,bukov15})
\begin{equation}
  \Uo (t_2, t_1) = \Uo_F (t_2) \, \re^{-\ri \Ho_F (t_2 - t_1) / \hbar} \Uo_F^{\dag} (t_1).
\end{equation}
Here, $\Uo_F (t)$ is the time-periodic micromotion operator and $\Ho_F$ is the effective 
Hamiltonian, which in addition to being stationary is also free from any parametric 
dependence on the choice of the initial and final instants of time. In this way the
effects of micromotion are clearly separated from the effective long-term dynamics.
The effective Hamiltonian can be systematically approximated in terms of series in the
inverse driving frequency (see Appendix~\ref{sec:chf} for details). In the present section, 
we focus on the single-particle properties and thus restrict our attention to a two-term 
approximation of the effective Hamiltonian
\begin{equation}
\label{eq:brak2}
\begin{split}
  \Ho_F^{[2]} &= \Ho_F^{(1)} + \Ho_F^{(2)} \\
  &= \Ho_0 + \frac{1}{\hbar\omega} \sum_{s=1}^{\infty}
  \frac{1}{s} \big[ \Ho_{s}, \Ho_{-s} \big].
\end{split}
\end{equation}
Third-order terms contained in $\Ho_F^{(3)}$ (\ref{eq:hf3}) are relevant for the coupling 
between kinetics and interactions, and will be included in the subsequent Sec.~\ref{sec:inter}.

\subsection{\myblue Single-particle spectra}

At each point $\bk$ in the Brillouin zone the Hamiltonian $\Ho_F^{[2]} (\bk)$ defines a 
two-level system, and therefore, is represented by a dot product of a three-dimensional 
Bloch vector $\mathbf{h} (\bk) = \{ h_x (\bk), h_y (\bk), h_z (\bk)\}$ and the vector 
of Pauli matrices $\{\sigma_x, \sigma_y, \sigma_z\}$:
\begin{equation}
  \Ho_F^{[2]} (\bk) = \mathbf{h} (\bk) \cdot \bm{\sigma}. 
\end{equation}
The lowest-order contribution to the effective Hamiltonian is given by the time-average 
of the driven Hamiltonian, or in other words, by its zeroth Fourier component, hence
\begin{equation}
  H_{F}^{(1)} 
  = -J \rJ_1 (\alpha) \sum_{\bk} \ba_{\bk} \ao_{\bk} G_1 (\bk) + \mathrm{h.c.}.
\end{equation}
This term contributes only to off-diagonal components $h_x (\bk)$ and $h_y (\bk)$. 
Obviously, the overall band width is modulated by the Bessel function $\rJ_1(\alpha)$, 
and the energy spectrum consists of two mirror-symmetric bands given by
\begin{equation}
  \varepsilon_{\pm} (\bk) = \pm J \left| \rJ_1 (\alpha) \right| \left|G_1(\bk)\right|.
\end{equation}
The analysis of the properties of the function $G_1 (\bk)$ presented in 
Appendix~\ref{app:fungus} reveals the presence of two Dirac points, where $h_x = h_y = 0$ 
and both bands touch in a conical fashion. One Dirac point is situated at the center of 
the Brillouin zone, 
\[
  \rD_1 : (k_1 = 0, k_2 = 0),
\]
and the other one is at a corner of the square-shaped Brillouin zone,
\[
  \rD_2 : \left(k_1 = \tfrac{1}{2}, k_2 = \tfrac{1}{2}\right).
\] 

\begin{figure}
\begin{center}
\includegraphics[width=84mm]{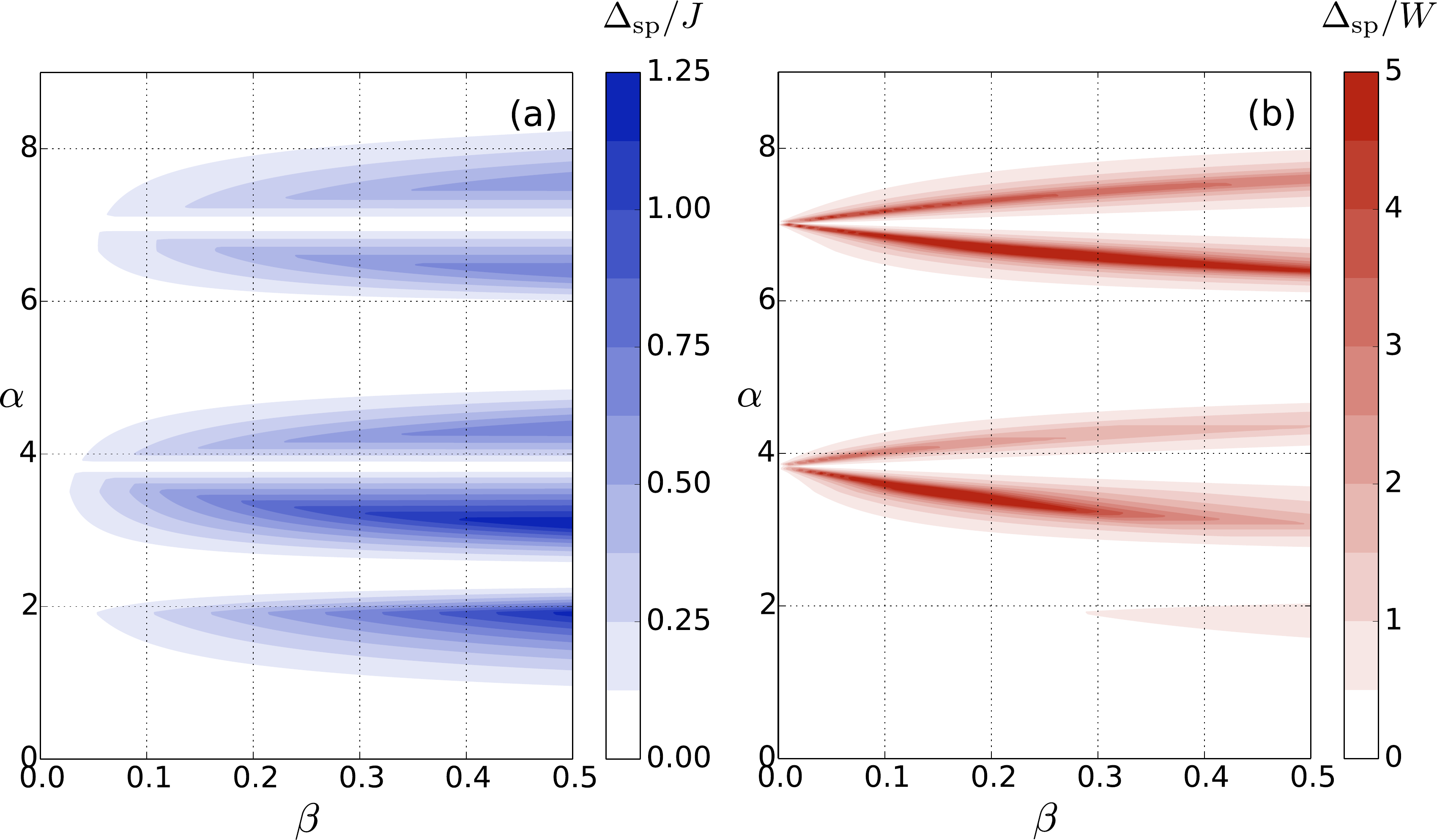}
\caption{\label{fig:spp} The single-particle band structure of the driven square lattice 
versus the scaled inverse driving frequency $\beta$ and the scaled driving strength 
$\alpha$. Panel (a) shows the width of the topological band gap $\Delta_{\textrm{sp}}$ 
(when present) measured in units of the hopping parameter $J$. In panel (b), the band 
gap is compared to the width of a single band thus defining the band flatness ratio
$\mathscr{F} = \Delta_{\textrm{sp}} / W$.}
\end{center}
\end{figure}

Proceeding to include the driving-induced second-order hopping transitions, we evaluate 
the commutators
\begin{equation}
\label{eq:sxt}
\begin{split}
  \big[ \Ho_s,\Ho_{-s} \big] &= J^2 \sum_{\bk} 
  \left( \aa_{\bk}\ao_{\bk} - \ba_{\bk} \bo_{\bk} \right)
  \Big\{ 
  \rJ_{s-1}^2 (\alpha) \left|G_{1-s} (\bk) \right|^2 \\
  &- \rJ_{1+s}^2 (\alpha) \left|G_{1+s} (\bk) \right|^2 
  \Big\}.
\end{split}
\end{equation}
These corrections are diagonal in the subband index and thus connect a given site to 
its next-nearest neighbors as well as further sites belonging to \emph{the same} 
sublattice. Note, that the mirror symmetry between the lower and upper energy bands is 
preserved and the second order correction $\Ho_F^{(2)} (\bk)$ has only $h_z (\bk)$ 
component. 

The Pauli-matrix representation makes the study of the topological nature of the bands 
straightforward: For the two energy bands to acquire nontrivial Chern numbers $\pm 1$, 
the Bloch vector $\mathbf{h} (\bk)$ must wrap around the full Bloch sphere 
\cite{alba11,sticlet12,goldman13} 
visiting its both poles. Consequently, to open a topological band gap $h_z (\bk)$ should 
have opposite signs at the two Dirac points. From Eq.~(\ref{eq:sxt}) we infer (see 
also Appendix \ref{app:fungus}) that at the Dirac points $\rD_{1|2}$, the contributions 
of the second-order expansion term $\Ho_F^{(2)}$ are given by an overall prefactor 
$16 J^2/\hbar\omega$ times the $\alpha$-dependent oscillatory factors; at $\rD_1$ 
\begin{subequations}
\begin{equation}
\label{eq:gap1}
  \rJ_0^2 (\alpha) - \left(\tfrac{1}{3} - \tfrac{1}{5}\right) \rJ_4^2 (\alpha)
  - \left(\tfrac{1}{7} - \tfrac{1}{9}\right) \rJ_8^2 (\alpha) - \ldots,
\end{equation}
and at $\rD_2$
\begin{equation}
\label{eq:gap2}
  - \left(1 - \tfrac{1}{3} \right) \rJ_2^2 (\alpha)
  - \left(\tfrac{1}{5} - \tfrac{1}{7}\right) \rJ_6^2 (\alpha) - \ldots.
\end{equation}
\end{subequations}
For most values of $\alpha$, the factors (\ref{eq:gap1}) and (\ref{eq:gap2}) have 
opposite signs that conspire with opposite chiralities of the Dirac points to
produce a topological band gap. Exceptions occur only in close vicinity of the zeros 
of $\rJ_0 (\alpha)$; here subleading terms of Eq.~(\ref{eq:gap1}) take over and the 
band structure features a trivial gap. The single-particle band structure generated 
by the two-term expansion $\Ho_F^{[2]}$ in Eq.~(\ref{eq:brak2}) is plotted in 
Fig.~\ref{fig:spp} as a function of the scaled inverse driving frequency
\begin{equation}
  \beta = \frac{J}{\hbar\omega},
\end{equation}
and the scaled driving strength 
\begin{equation}
  \alpha = \frac{Fd}{\hbar\omega}.
\end{equation}
In the left panel the shades of blue indicate the size of the \emph{topological} gap, 
if present, that separates the upper and lower energy bands characterized by the Chern 
indices $\pm 1$. To be fully precise here, the band gap $\Delta_{\textrm{sp}}$ is defined 
as the \emph{global gap}, that is
\begin{equation}
\begin{split}
  \Delta_{\text{sp}} &= \operatorname*{min}_{\bk \in \text{BZ}} \varepsilon_{+} (\bk) 
   - \operatorname*{max}_{\bk \in \text{BZ}} \varepsilon_{-} (\bk).
\end{split}
\end{equation}
Owing to the mirror-symmetry of the energy bands, the band gap is direct. The band
gap is of the order $J^2 / \hbar\omega$ and, thus, vanishes for small $\beta$. This
indicates that our scheme belongs to those working at intermediate rather than large
driving frequencies. The right 
panel of Fig.~\ref{fig:spp} shows the band flatness defined as the ratio of the 
topological band gap to the overall width of a single band, viz.
\begin{equation*}
  \mathscr{F} = \frac{\Delta_{\text{sp}}}{W}, \qquad 
  W = \operatorname*{max}_{\bk \in \text{BZ}} \varepsilon_{+} (\bk) 
  - \operatorname*{min}_{\bk \in \text{BZ}} \varepsilon_{+} (\bk).
\end{equation*}
This quantity is more relevant in the context of stabilization of the fractional Chern 
insulator phases, ideally requiring topological band gaps exceeding (or at least 
comparable to \cite{grushin15stability,kourtis14}) the interaction strengths which, in 
their own turn, must dominate the band width. We see that robust flatness ratios in 
excess of $4$ can be reached.

\section{Interplay of micromotion and interactions}
\label{sec:inter}

In order to gain insight into the effect of coupling between particle tunneling and 
interaction events that appear in the third-order of the effective Hamiltonian 
(\ref{eq:hf3}), we consider here a generic situation described by a driven kinetic 
Hamiltonian
\begin{equation}
\label{eq:trtfr}
  \Ho (t) = - J \sum_{\la ij \ra} \re^{\ri \theta_{ij} (t)} \, \aa_i \ao_j,
\end{equation}
with Peierls phases of the form (\ref{eq:aten})
\begin{align}
  \theta_{ij} &= \alpha \sin (\omega t - \ph_{ij}) + s_{ij} \, \omega t.
\end{align}
In contrast to the $\bk$-space computational approach taken in the previous 
Sec.~\ref{sec:model}, 
we now write the Fourier components of the Hamiltonian (\ref{eq:trtfr}) in the real 
space
\begin{equation}
  \Ho_s = -J \sum_{\la ij \ra} \rJ (s-s_{ij} | \alpha) \,
  \re^{-\ri (s - s_{ij})\ph_{ij} } \aa_i \ao_j.
\end{equation}
Here, in order to avoid the clumsiness of double subscripts we introduced an in-line 
notation for the Bessel functions $\rJ (s | \alpha) \equiv \rJ_s (\alpha)$.
Let us next assume that interactions between particles are bosonic and on-site, i.~e.,
\begin{equation}
  \Ho_{\inter} = \frac{U}{2} \sum_{i} \aa_{i} \aa_{i} \ao_{i} \ao_{i},
\end{equation}
and evaluate the nested commutator that defines the third-order contribution to the 
effective Hamiltonian
\begin{widetext}
\begin{equation}
\begin{split}
  \big[ \big[ \Ho_{\inter}, \Ho_s \big], \Ho_{-s} \big] 
  &= \frac{U J^2}{2} \sum_i \sum_{\la jk \ra} \sum_{\la \ell m \ra} 
    \big[ \big[ \aa_i \aa_i \ao_i \ao_i, \aa_j \ao_k \big], \aa_{\ell} \ao_m \big] \\
  &\times \rJ (s-s_{jk} | \alpha) \rJ (-s-s_{\ell m} | \alpha)
    \, \re^{-\ri (s-s_{jk}) \ph_{jk}} \, \re^{-\ri(-s-s_{\ell m})\ph_{\ell m}}.
\end{split}
\end{equation}
Focusing on the basic structural element in this expression we obtain
\begin{equation}
\begin{split}
\label{eq:threelines}
  \big[ \big[ \aa_i \aa_i \ao_i \ao_i, \aa_j \ao_k \big], \aa_{\ell} \ao_m \big] 
  &= 2 \delta_{ij} \delta_{i\ell} \, \aa_i \aa_i \ao_k \ao_m 
   + 2 \delta_{ik} \delta_{im} \, \aa_j \aa_{\ell} \ao_i \ao_i \\
  &- 4 \delta_{ij} \delta_{im} \, \aa_{\ell} \aa_i \ao_i \ao_k 
   - 4 \delta_{ik} \delta_{i\ell} \, \aa_j \aa_i \ao_i \ao_m \\ 
  &+ 2 \delta_{ij} \delta_{k\ell} \, \aa_i \aa_i \ao_i \ao_m 
   + 2 \delta_{ik} \delta_{jm} \, \aa_{\ell} \aa_i \ao_i \ao_i.
\end{split}
\end{equation}
\end{widetext}

The individual terms in this result admit a clear physical interpretation (see also 
Refs.~\onlinecite{eckardt15,anisimovas15}). 
We easily recognize that the first line of Eq.~(\ref{eq:threelines}) lists 
pair hopping events: Two particles are removed from site $i$ and placed onto two of 
its nearest-neighboring sites (either the same site or two distinct sites). The 
conjugate version allows two particles to hop onto site $i$ from its two neighboring 
sites. The second line 
of Eq.~(\ref{eq:threelines}) introduces density-assisted hopping events between two 
nearest neighbors of site $i$ using site $i$ as the intermediate stop. If the two
nearest neighbors coincide, however, these terms transform into the ordinary 
density-density interactions between nearest neighbors. Finally, the third and last
line lists events where a particle leaves a given site $i$ and travels to another site 
reachable by two nearest-neighbor transitions. If the origin and the destination coincide,
these contributions degenerate into the ordinary on-site repulsion. As we will see
shortly, these contributions are negative in the sense that the original on-site 
repulsion energy $U$ is effectively reduced.

Summarizing the above observations, we note that various contributions induced by 
combining kinetic and interaction events can be categorized into three classes:
(i) terms contributing to the modification of the on-site repulsion energy $U$,
(ii) terms leading to the introduction of hitherto absent density-density 
interaction between nearest neighbor sites, and (iii) more exotic density-assisted 
and pair tunneling events. It is quite remarkable that the two former contributions 
obey a constraint that can be interpreted as a conservative partial spread of on-site
interactions onto the nearest-neighbor interactions.

In order to demonstrate the advertised result, we review the commutators
in Eq.~(\ref{eq:threelines}) and collect only the terms that contribute to (the 
reduction of) on-site interactions and obtain
\begin{equation}
\label{eq:commr}
\begin{split}
  &\big[ \big[ \aa_i \aa_i \ao_i \ao_i, \aa_j \ao_k \big], 
    \aa_{\ell} \ao_m \big]_{\textrm{R}}\\
  &\qquad = 2 \left( \delta_{ij} \delta_{k\ell}\delta_{im} 
  +  \delta_{ik} \delta_{jm} \delta_{i\ell} \right)
   \aa_i \aa_i \ao_i \ao_i.
\end{split}
\end{equation}
Here, the subscript $\textrm{R}$ refers to \emph{renormalization} of the original 
on-site repulsion energy. Likewise, filtering out terms that contribute to the
appearance of a nearest-neighbor repulsion we obtain
\begin{equation}
\label{eq:commn}
\begin{split}
  &\big[ \big[ \aa_i \aa_i \ao_i \ao_i, \aa_j \ao_k \big], 
  \aa_{\ell} \ao_m \big]_{\text{N}} \\
  &\qquad = -4 \left( \delta_{ij} \delta_{k\ell} \delta_{im}\, \no_i \no_{\ell} 
    +  \delta_{ik} \delta_{jm} \delta_{i\ell} \no_i \no_m \right),
\end{split}
\end{equation}
with the subscript $\textrm{N}$ serving as a mnemonic for interactions with neighboring 
sites.

Close resemblance of the results in basic commutators (\ref{eq:commr}) and 
(\ref{eq:commn}) survives the lattice summations, and leads to the correction terms 
in the effective Hamiltonian
\begin{subequations}
\begin{align}
  \label{eq:hfr}
  \left[\Ho_F^{(3)} \right]_R &= 
    \frac{\Delta U}{2} 
    \sum_i \aa_i \aa_i \ao_i \ao_i, \\
  \label{eq:hfn}
  \left[\Ho_F^{(3)} \right]_N &= 
    \frac{\Delta V}{2} 
    \sum_{\la ij \ra} \no_i \no_j. 
\end{align}
\end{subequations}
Here
\begin{equation}
  \Delta U = - z f U, \quad\text{and}\quad
  \Delta V = 2 f U,
\end{equation}
with $z$ denoting the number of nearest neighbors on the lattice (the coordination number)
and the renormalizing factor reads
\begin{equation}
  f = \frac{2 J^2}{(\hbar\omega)^2} 
  \sum_{s=1}^{\infty} \frac{\rJ_{s-1}^2 (\alpha) + \rJ_{s+1}^2 (\alpha)}{s^2}.
\end{equation}
In conclusion, lattice shaking leads to a redistribution of the on-site interaction
energy whereby it is partially spread onto nearest-neighbor interactions with the 
constraint
\begin{equation}
\label{eq:sumrule}
  \Delta U + \frac{z}{2} \Delta V = 0,
\end{equation}
relating the changes in the respective energies. The appearance of the factor $z/2$ 
is easy to understand: Each lattice site serves as an endpoint of $z$ nearest-neighbor 
links but due to a double counting of the links only half of them belong to a given site.

\section{Numerical phase diagrams}
\label{sec:numpy}

Let us now proceed to the numerical illustrations of the above general discussion.
Starting from the single-particle band structure supported by chiral $\pi$-flux model of 
Sec.~\ref{sec:model} we take into account also bosonic on-site inter-particle
interactions and perform exact diagonalizations on a finite lattice of
$4 \times 4$ elementary cells containing $32$ sites with periodic or twisted
boundary conditions. From the single-particle point of view, one thus realizes two 
energy bands supported on a discrete grid of $4 \times 4$ points in the Brillouin zone.
Introducing $N_p = 8$ interacting particles we fill the lower band at the filling factor
$\nu = \tfrac{1}{2}$ where bosonic fractional Chern insulator states are expected to 
form. The numerical procedure of identifying the fractional states and computing
the many-body topological band gap --- that separates the ground state manifold from the 
rest of the spectrum --- are described in detail elsewhere \cite{regnault11,anisimovas15}. 
To summarize briefly,
the exact diagonalizations are repeated multiple times: One samples over the many-body
Brillouin zone spanned by the auxiliary fluxes $(\gamma_1, \gamma_2)$ that define the 
twisted boundary conditions in the two directions. The obtained data is used to extract 
the topological many-body gap as the minimum separation between the ground state
manifold and the excited states. Also, the behavior of the states in response to the 
changing auxiliary fluxes confirms their fractional nature. The numerical procedure
is simplified by: (i) quasimomentum conservation that allows to carry out computations
separately at each individual point in the reciprocal space, and (ii) the existing 
predictions for the quasimomentum sectors at which fractional states 
will be formed \cite{regnault11}.

The purpose of the numerical simulations is twofold. Firstly, we must check whether the 
presented realization of the chiral $\pi$-flux model can sustain fractional Chern 
insulator phases. The single-particle band structure is promising in terms of the 
presence of a topological gap and significant band flatness. Therefore it 
is interesting to see if topological many-body states can be stabilized and if they 
withstand the impact of micromotion which was shown to be largely detrimental, 
in particular at lower driving frequencies \cite{anisimovas15}. 
Secondly, the analysis of the previous 
Sec.~\ref{sec:inter} allows us to separate the distinct constituent components of the 
interplay between micromotion and interactions. It is therefore very relevant to ask 
what role is played by the reduction of the on-site interaction strength, the 
appearance of nearest-neighbor repulsion, and finally, the density-assisted 
tunneling events.

\begin{figure}
\begin{center}
\includegraphics[width=84mm]{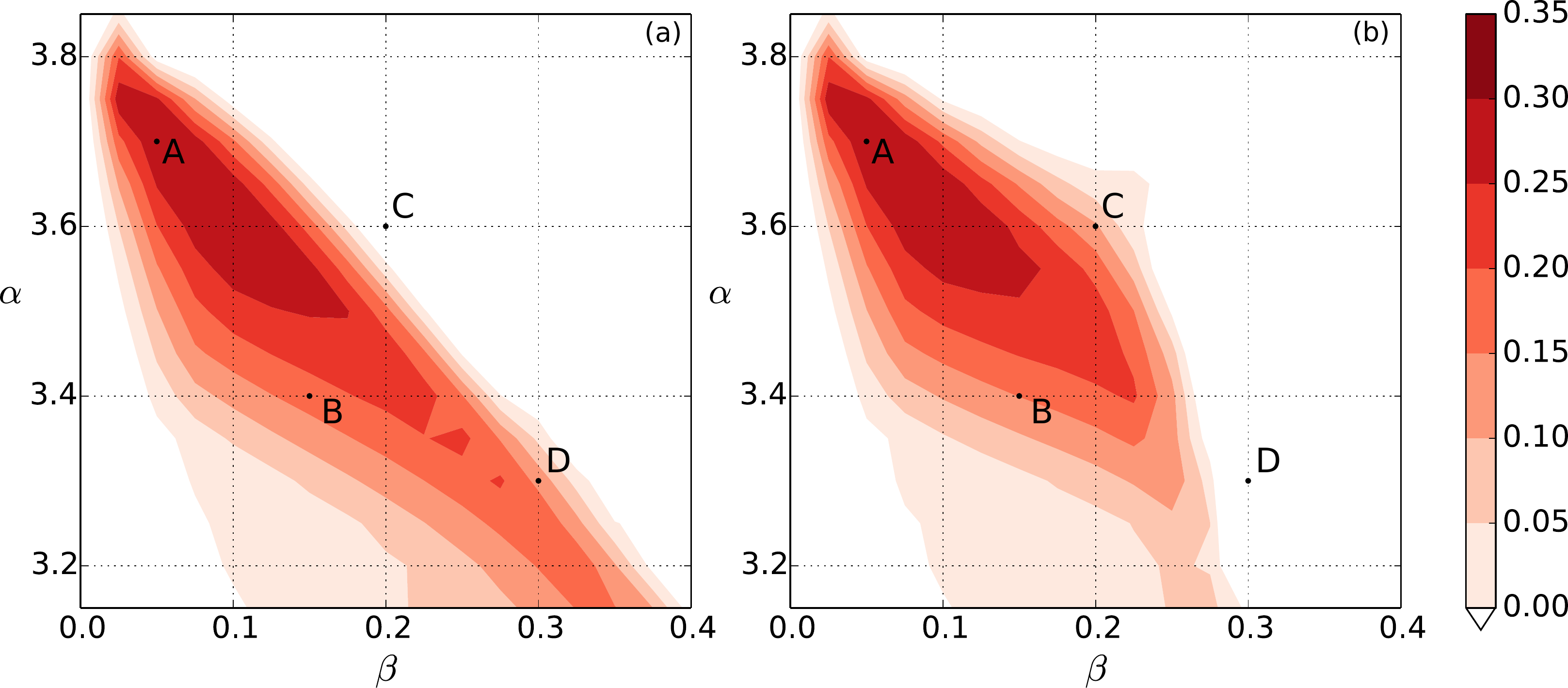}
\caption{\label{fig:phd} Many-body topological gap measured in units $J$ for eight bosons 
moving on a lattice $4 \times 4$ two-site unit cells: (a) micromotion omitted, 
(b) micromotion included. The points labeled with the letters A, B, C, and D correspond 
to four typical behavioral patterns further analyzed in Fig.~\ref{fig:abcd}.}
\end{center}
\end{figure}

We start with the issue of stability of fractional states, and show in 
Fig.~\ref{fig:phd} the phase diagrams of
eight bosons moving on a $4 \times 4$ lattice. Here, the dependence of the 
topological many-body gap is plotted as a function of the scaled shaking strength
$\alpha = Fd / \hbar\omega$ and scaled inverse frequency $\beta = J / \hbar\omega$
varying in the region of the largest band flatness detected
in the single-particle simulations [see Fig.~\ref{fig:spp}~(a)]. The left panel (a)
refers to the case where micromotion is neglected, that is, the third-order
term $\Ho_F^{(3)}$ is not included. In contrast, the right panel (b) presents the 
results obtained with micromotion taken into account. The colored regions correspond
to the presence of a topological many-body gap, with the intensity encoding the size 
of the gap. The visible change of the shape indicates that micromotion has a
significant impact on the stability of the fractional Chern insulator phase. 
The many-body gap closes for small $\beta$, since the single-particle gap closes for
too large driving frequencies. However, it also closes for large values of $\beta$, 
that is, too slow shaking. To aid further analysis, we define four representative 
reference points A, B, C, and D seen in the phase diagrams of Fig.~\ref{fig:phd}. 
At points A and B, micromotion seems to have no perceptible impact on the size of 
the topological band gap. At point C, the stability region has a shoulder where 
micromotion seems to enhance the fractional phase. Finally, at point D micromotion 
has a clearly detrimental effect. Here, the fractional Chern insulator phase is 
strongly suppressed.

\begin{figure}
\begin{center}
\includegraphics[width=84mm]{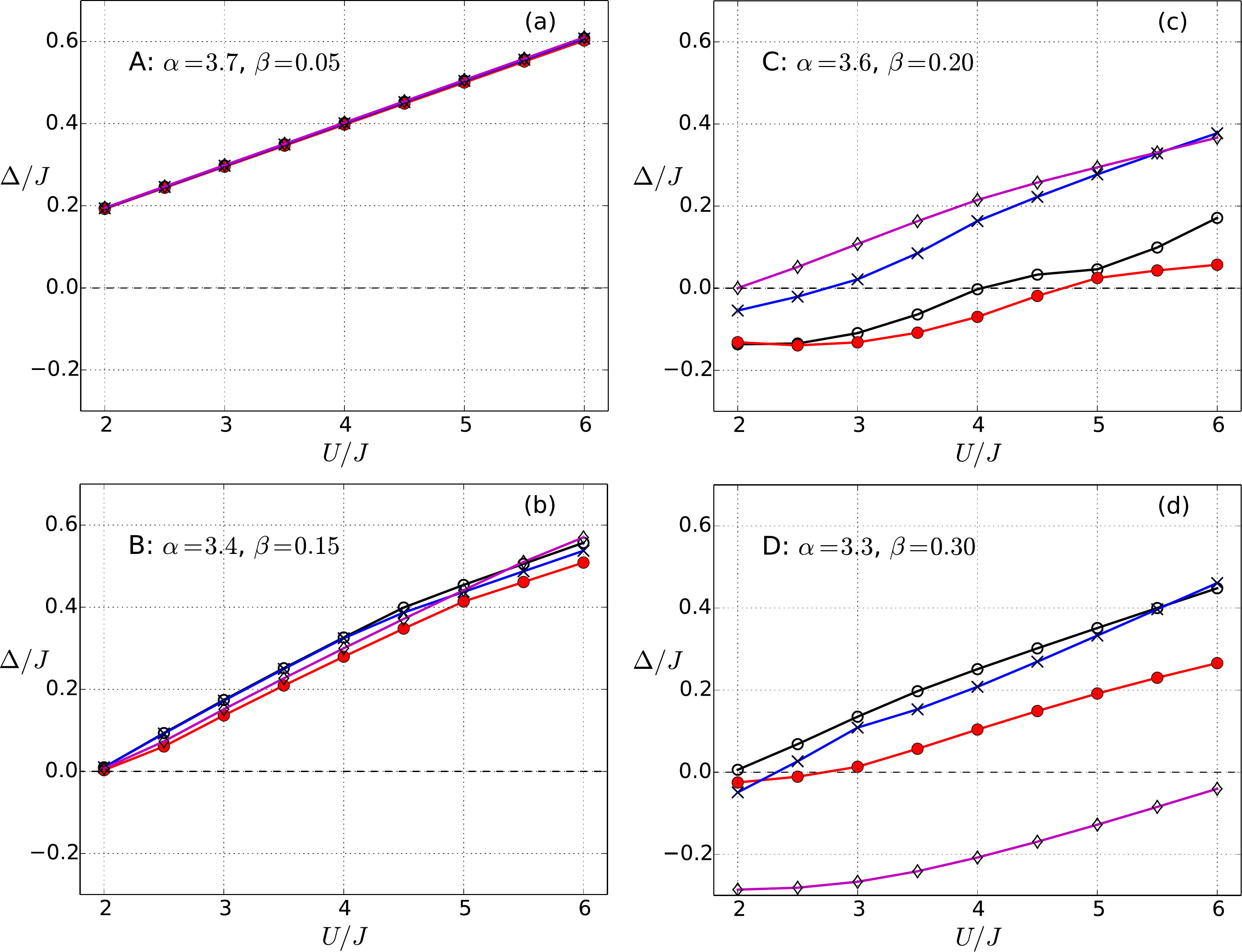}
\caption{\label{fig:abcd}Many-body gap for eight bosons moving on a lattice $4 \times 4$
at four typical parameter sets A, B, C, and D marked in Fig.~\ref{fig:phd}. Black lines 
and empty circle markers: micromotion completely neglected, red lines with full circle 
markers: only reduction of on-site repulsion strength taken into account, blue lines with 
crosses: on-site repulsion spread onto neighboring sites, purple lines with rhombus-shaped 
markers: all third-order corrections included.}
\end{center}
\end{figure}

The four reference locations on the phase diagram A, B, C, D exemplify four typical
patterns observed in the interplay of micromotion and interactions and shown in 
Fig.~\ref{fig:abcd}. The four panels of this figure correspond to the four points
and show the growth of the many-body gap as a function of the bare on-site repulsion
strength $U$. The black lines connecting empty circles show the results obtained
in the absence of micromotion effects, that is, when $\Ho_F^{(3)}$ is not taken into account.
The red lines connecting data points marked by full circles correspond to the crudest 
approximation of the effects of micromotion on interactions: Here, only the on-site
repulsion strength is reduced by including the correcting term (\ref{eq:hfr}).
As expected, this reduction leads to smaller many-body gaps as in all four panels 
the red line lies below the black one. As we have shown in the previous subsection, 
coupling between micromotion and interactions does not simply reduce the on-site
repulsion energy but rather spreads it onto nearest-neighboring sites. Therefore,
in our plot we include also the case when both these effects [i.~e., both corrections
(\ref{eq:hfr}) and (\ref{eq:hfn})] are taken into account. The behavior of the many-body
gap is now shown by the blue lines connecting cross-shaped markers. Evidently, the
proper inclusion of micromotion-induced nearest-neighbor interactions generally has 
a significant stabilizing effect. Finally, the purple lines drawn over rhombus-shaped
markers correspond the calculations that fully take the third-order correction
into account. Thus, comparing the relative positions of purple and blue lines one
may estimate the role of the density-assisted processes included at this last stage. 

Proceeding to the analysis of the four typical patterns in panels (a)--(d) of
Fig.~\ref{fig:abcd}, we begin with the reference point A in the top left corner of the 
shown phase diagram. Here, the fractional Chern insulator state is robust and the
effect of third-order processes is minor due to the very small value of $\beta = 0.05$
which corresponds to the high-frequency limit. Consequently, the four lines corresponding
to the different levels of approximation nearly coincide in Fig.~\ref{fig:abcd}~(a). 
Moving along the line connecting point A to point 
B, one stays in the region characterized by large single-particle band flatness and
robust fractional Chern insulator phase. However, the driving frequency is progressively 
lowered and the parameter $\beta$ reaches the value of $0.15$ at point B. Here, the 
four lines seen in Fig.~\ref{fig:abcd}~(b) start to diverge indicating that individual
micromotion-related contributions have either a negative (reduction of on-site repulsion), 
positive (nearest-neighbor interactions) or varying (density-assisted hopping events) effect.
However, the individual contributions largely cancel out and there is only a minor 
modification to the stability of the fractional phase.

Next, it is interesting to look at point C where the phase diagram indicates that 
micromotion-induced interactions may have a positive role on the topological many-body
band gap. The results shown in Fig.~\ref{fig:abcd}~(c) show that this enhanced stability 
is mainly due to the positive impact of induced nearest-neighbor interactions. Finally,
point D corresponds to relatively slow driving where micromotion has a very strong
and evidently negative impact on the stability of the fractional Chern insulator. Our
results shown in Fig.~\ref{fig:abcd}~(d) reveal that the destruction of the fractional 
phase is mainly due to rapidly growing negative contribution from the density-assisted 
hopping events.

\section{Summary and conclusions}
\label{sec:sum}

To summarize, we have proposed a scheme for the realization of a Floquet topological 
band structure in a circularly shaken square lattice. Moreover, we have shown that 
in a driven system of interacting particles the presence of the real-space micromotion 
couples to the on-site inter-particle interactions and leads to the appearance of 
additional interaction terms in the effective Hamiltonian that allow for a physically 
transparent interpretation and classification. One part of the micromotion-induced 
contributions produces the effect of ``smearing out'' the on-site repulsion onto the 
nearest neighbors. The original (bare) on-site interaction energy is diminished and 
the missing portion is distributed over the links to nearest neighbors in a way that 
satisfies a strict constraint (\ref{eq:sumrule}). Another part of the effects of 
micromotion lead to the appearance of density-assisted hopping terms. Applying our 
description to the bosonic fractional Chern insulator states at the filling factor 
$\nu = 1/2$, we elucidate the role of the individual contributions on the stability 
of the fractional phase. It turns out that the fractional Chern insulators are largely
destabilized by the density-assisted hopping events when their contribution becomes 
sufficiently large at lower driving frequencies. Note that also heating effects related 
to the creation of collective excitations (not captured by the high-frequency expansion) 
will become more and more important with decreasing driving frequency \cite{eckardt15}. 
This suggests that the conditions for the preparation of a Floquet fractional Chern 
insulator are most favorable in the regime of rather large frequencies 
($\hbar\omega \sim 10J$).

\acknowledgments

This work was supported by the Lithuanian Research Council under the Grant No.~MIP-086/2015. 
We thank Gediminas Juzeli\={u}nas and Brandon M.~Anderson for insightful discussions.

\bibliography{mint,experiments,fti,fractional,driven,optical,topins}

\appendix

\section{\myblue Artificial gauge structures in driven lattices}
\label{app:gauge}

A generic lattice model is defined by a set of lattice sites $i$ connected pairwise by 
a complementary set of connecting links. In our notation, $\la ij \ra$ denotes a 
\emph{directed} link originating at site $j$ and running to site $i$. Thus, a purely 
kinetic Hamiltonian of a static (undriven) lattice reads
\begin{equation}
  \Ho_{\text{st}} = - \sum_{\la ij \ra} J_{ij} \aa_i \ao_j.
\end{equation}
Here, $J_{ij}$ denotes the bare transition parameter, and $\ao^{(\dag)}_i$ are the 
standard annihilation (creation) operators defined on site $i$. Even though in standard 
situations $J_{ij}$ are real positive quantities, this model can be endowed with an 
artificial gauge structure by means of a time-periodic driving. In the present work, 
in addition to the lattice driving we also rely on the presence of staggered on-site 
potential shifts. Both effects are captured together by a potential-energy term
\begin{equation}
\label{eq:drivpot}
  \Vo (t) = \sum_{i} v_i (t) \aa_i \ao_i,
\end{equation}
written as a sum over lattice sites and featuring time- and coordinate-dependent 
on-site potentials of the form
\begin{equation}
  v_i (t) = - \bF (t) \cdot \br_i + s_i \hbar\omega.
\end{equation}
Here, $\br_i$ is the position vector of the site $i$, $\bF(t)$ is a time-periodic 
driving force oscillating with frequency $\omega$, and the static on-site energy shifts 
are characterized by integer factors $s_i$ times the energy quantum $\hbar\omega$. The 
potential term (\ref{eq:drivpot}) can be eliminated from the Hamiltonian by means of a 
unitary transformation
\begin{equation}
  \Uo (t) = \prod_i \Uo_i (t), \quad
  \Uo_i (t) = \exp \left[- \ri\, \chi_i (t) \aa_i \ao_i \right],
\end{equation}
written as a product of independent operators acting on a single site and fulfilling 
the cancellation condition
\begin{equation}
  \Ua_i (t) v_i (t) \aa_i \ao_i \Uo_i (t) - \ri\hbar \Ua_i (t) \p_t \Uo_i (t) = 0.
\end{equation}
The phases 
$\chi_i (t)$ are obtained through a straightforward time integration
\begin{equation}
\begin{split}
  \chi_i (t) &= \frac{1}{\hbar} \int_0^t \! dt' \, v_i (t') + \gamma_i \\
  &= - \frac{1}{\hbar} \int_0^t \! dt' \, \bF (t') \cdot \br_i + s_i \omega t
  + \gamma_i,
\end{split}
\end{equation}
with integration constants $\gamma_i$ reflecting the gauge freedom. The transformed
Hamiltonian is purely kinetic
\begin{equation}
\label{eq:nonint}
  \Ho (t) = \Uo^{\dag} (t) \Ho_{\text{st}} \Uo (t)  
  = - \sum_{\la ij \ra} J_{ij} \, \re^{\ri \theta_{ij} (t)} \aa_i \ao_j,
\end{equation}
and features the Peierls phases
\begin{equation}
\begin{split}
  \theta_{ij} (t) &= \chi_i (t) - \chi_j (t)  \\
  &= - \frac{1}{\hbar} \int_0^t \! dt' \, \bF (t') \cdot \br_{ij}
  + s_{ij} \, \omega t + \gamma_{ij}.
\end{split}
\end{equation}
Here, $\br_{ij} = \br_i - \br_j$, and likewise $s_{ij} = s_i - s_j$ and 
$\gamma_{ij} = \gamma_i - \gamma_j$.

In this work, we assume a circular driving protocol
\begin{equation}
  \bF (t) = - \hat{\bm e}_x F \cos (\omega t + \phi_0) 
  - \hat{\bm e}_y F \sin (\omega t + \phi_0),
\end{equation}
equal hopping distances $|\br_{ij}| \equiv d$ and bare hopping parameters 
$J_{ij} \equiv J$, and define the polar angles $\ph_{ij}$
measuring the direction of the vector $\br_{ij}$ with respect to the $x$ axis. Under 
these assumptions, the Peierls phases read
\begin{equation}
\label{eq:aten}
  \theta_{ij} (t) = \alpha \sin \left(\omega t - \ph_{ij} + \phi_0 \right) + 
  s_{ij} \, \omega t, 
\end{equation}
with the dimensionless shaking strength $\alpha = Fd / \hbar\omega$.

\section{\myblue High-frequency expansion of the effective Hamiltonian}
\label{sec:chf}

The long-term dynamics generated by a time-periodic Hamiltonian is given by a 
time-independent effective Hamiltonian, and one's ability to deliberately engineer its 
properties is known as the Floquet engineering. A standard approach to the construction 
of the effective 
Hamiltonian employs a series expansion in the powers of the inverse driving frequency. 
The successive terms in the series are conveniently expressed in terms of operator-valued 
Fourier components $\Ho_s$ of the driven Hamiltonian. Therefore we Fourier analyze the
kinetic Hamiltonian
\begin{equation}
  \Ho (t) = \sum_{s=-\infty}^{\infty}  \Ho_s \,\re^{\ri s\omega t}.
\end{equation}
Specializing to static density-density interactions between particles we add them 
to the zeroth (static) Fourier component
\begin{equation}
  \Ho_0 \quad\to\quad \Ho_0 + \Ho_{\inter}
\end{equation}
The leading terms of the high-frequency expansion of the effective Hamiltonian 
\cite{goldman14,eckardt15,goldman15resonant,bukov15,itin15,mikami15}
\begin{equation}
  \Ho_F = \Ho_F^{(1)} + \Ho_F^{(2)} + \Ho_F^{(3)} + \cdots
\end{equation}
read 
\begin{subequations}
\begin{align}
  \label{eq:hf1}
  \Ho_F^{(1)} &= \Ho_0 + \Ho_{\inter}, \\
  \label{eq:hf2}  
  \Ho_F^{(2)} &= \frac{1}{\hbar\omega}
    \sum_{s=1}^{\infty} \frac{1}{s} \big[\Ho_s,\Ho_{-s}\big], \\
  \label{eq:hf3}
  \Ho_{F}^{(3)} &= - \frac{1}{2(\hbar\omega)^2} \sum_{s=1}^{\infty} \frac{1}{s^2}
  \big[ \big[ \Ho_0 + \Ho_{\inter}, \Ho_s \big], \Ho_{-s} \big] + \mathrm{h.c.}. 
\end{align}
\end{subequations}
The commutator structure of the above expansion provides a clear physical interpretation
of the successive contributions. Going beyond the high-frequency limit given by the
time-averaged driven Hamiltonian $H_F^{(1)}$, in $\Ho_F^{(2)}$ one encounters combined 
hopping events that introduce next-neighbor transitions and help open a topological band 
gap in certain models, including also 
the example presented in this contribution. The physics is further enriched by the 
third order contribution $\Ho_F^{(3)}$, which can be separated into purely kinetic 
processes \cite{mikami15} given by $\big[ \big[ \Ho_0, \Ho_s \big], \Ho_{-s} \big]$ 
and an interplay between the real-space micromotion and interactions encoded in 
$\big[ \big[\Ho_{\inter}, \Ho_s \big], \Ho_{-s} \big]$. In this work we focus
on the latter effect and will not take the purely kinetic contribution into account. 
This is justified by two reasons.
Firstly, the combined kinetic processes, being proportional to $\omega^{-2}$, are weak.
On the other hand, the terms describing the interplay of micromotion and interactions
are in addition scaled by the interaction strength, which in most situations relating
to the formation of nontrivial phases must be large. Another reason pertains to the
physical motivation -- it is our primary goal to study the effects created by 
combining tunneling and interaction processes.

\section{\myblue Analysis of the auxiliary function $G$}
\label{app:fungus}

To aid the analysis of the single-particle spectra, it is worthwhile to record some 
properties of the auxiliary function $G(\bk)$ defined by Eq.~(\ref{eq:driven3}). Let
us express the quasimomentum vector $\bk$ in terms of its components along the 
elementary reciprocal lattice vectors
\begin{equation}
  \bk = k_1 \bmb_1 + k_2 \bmb_2, \quad\text{with}\quad
  \bma_i \cdot \bmb_j = 2\pi\delta_{ij},
\end{equation}
so that $\bk \cdot \bma_j = 2\pi k_j$, and we obtain for the absolute value of the 
$G$-function 
\begin{equation}
\begin{split}
  \left| G_s (k_1, k_2)\right|^2 
  &= 4 
    + 4 \cos \left(\frac{s\pi}{2}\right) \left(\cos 2\pi k_1 + \cos 2\pi k_2 \right) \\
  &+ 4 (-1)^s \cos 2\pi k_1 \cos 2\pi k_2.
\end{split}
\end{equation}
Thus, for odd values of $s$ we have
\begin{equation}
\begin{split}
  \left| G_{s\text{ odd}} (k_1, k_2) \right|  
  & = 2\sqrt{1 - \cos 2\pi k_1 \cos 2\pi k_2} 
\end{split}
\end{equation}
This function vanishes at two points within a Brillouin zone, namely, 
the center $\rD_1 : (k_1 = 0, k_2 = 0)$, and the corner 
$\rD_2 : \left(k_1 = \tfrac{1}{2}, k_2 = \tfrac{1}{2}\right)$ of the square-shaped
Brillouin zone. These are the two inequivalent Dirac points defined by the time-averaged
driven Hamiltonian, hence the notation $\rD_{1|2}$. 

Turning now to the $G$-functions of even indices, we observe that they separate into 
two classes
\begin{equation}  
\begin{split}
    \left| G_{s\text{ even}} (k_1, k_2)\right|^2 
  &= 4 \left(1 \pm \cos 2\pi k_1\right) \left(1 \pm \cos 2\pi k_2\right),
\end{split}
\end{equation}
depending on the index $s$ being (upper sign $+$) or not being (lower sign $-$) a 
multiple of four; thus
\begin{subequations}
\begin{align}
\label{eq:}
  |G_{4n} (k_1, k_2)|^2 &= 16 \, \cos^2 \pi k_1 \cos^2 \pi k_2,\\
  |G_{4n+2} (k_1, k_2)|^2 &= 16 \, \sin^2 \pi k_1 \sin^2 \pi k_2.
\end{align}
\end{subequations}
The values of these functions attained at the Dirac points are responsible for 
opening band gaps due to second-order hoppings. Thus we record (for an integer $n$)
at $\rD_1$:
\begin{equation}
  \left| G_{4n} (0, 0)\right|^2 = 16, 
  \qquad\left| G_{4n+2} (0, 0)\right|^2 = 0, 
\end{equation}
and at $\rD_2$:
\begin{equation}
  \left| G_{4n} \left(\tfrac{1}{2}, \tfrac{1}{2} \right) \right|^2 = 0, 
  \qquad\left| G_{4n+2} \left(\tfrac{1}{2}, \tfrac{1}{2} \right) \right|^2 = 16. 
\end{equation}
When summed over the Fourier components, these numerical values lead to the band gap
sizes given in Eq.~(\ref{eq:gap1}) and Eq.~(\ref{eq:gap2}) of the main text.

\end{document}